\documentclass[aps,prx,twocolumn,floats,showpacs,superscriptaddress,nofootinbib,nolongbibliography]{revtex4-2}
\usepackage{graphicx,epsfig}
\usepackage{times,bbm}
\usepackage{graphics,dcolumn,bm,float}
\usepackage{amssymb,amsmath,rotate,color,amsfonts}
\usepackage[title,titletoc,toc]{appendix}
\usepackage{mathtools}
\usepackage{booktabs}
\usepackage{enumitem}
\usepackage{tcolorbox}
\usepackage[pagebackref=false,colorlinks,linkcolor=magenta,citecolor=blue,urlcolor=magenta]{hyperref}

\usepackage{wrapfig}
\usepackage{lipsum}
\usepackage{mwe}
\usepackage[mathlines]{lineno}
\usepackage[mathscr]{euscript}
\usepackage{hyperref}
\usepackage{breqn}
\usepackage{bbold}
\usepackage{pgfplots}
\usepackage{float}
\usepackage{tkz-euclide}
\usepackage{braket}
\usepackage{physics}
\usepackage[caption=false]{subfig}
\usepackage[export]{adjustbox}
\usepackage{tikz}
\usepackage{slashed}
\usepackage{bm}
\usetikzlibrary{through,calc}
\usetikzlibrary{positioning}

\usepackage{subfig}
\usepackage{graphicx}
\usepackage{dcolumn}
\usepackage{bm}
\usepackage{comment} 

\newcommand{\be}{\begin{equation}}
\newcommand{\ee}{\end{equation}}
\newcommand{\ba}{\begin{align}}
\newcommand{\ea}{\end{align}}

\newcommand{\bea}{\begin{eqnarray}}
\newcommand{\eea}{\end{eqnarray}}

\newcommand{\bmt}{\left[\begin{matrix}}
\newcommand{\emt}{\end{matrix}\right]}

\begin{document}
\preprint{}
\title{Valley-Enabled Intrinsic Dresselhaus Spin-Orbit Coupling in Silicon}

\author{J. L. P. Steinschuld}
\affiliation{2nd Institute of Physics C$,$ RWTH Aachen University$,$ 52074 Aachen$,$ Germany}
\author{H. J. Bluhm}
\affiliation{2nd Institute of Physics C$,$ RWTH Aachen University$,$ 52074 Aachen$,$ Germany}
\author{L. R. Schreiber}
\affiliation{2nd Institute of Physics C$,$ RWTH Aachen University$,$ 52074 Aachen$,$ Germany}
\author{S. A. Jafari}
\affiliation{2nd Institute of Physics C$,$ RWTH Aachen University$,$ 52074 Aachen$,$ Germany}
\email{akbar.jafari@rwth-aachen.de}
\date{\today}

\begin{abstract}
We develop a symmetry-based theory of spin-orbit-valley coupling in silicon that reveals an intrinsic source of Dresselhaus spin-orbit coupling independent of interfaces or external electric fields. Treating the valley degree of freedom as a symmetry-carrying quantum degree of freedom, we show that the Dresselhaus interaction is necessarily valley off-diagonal and that a bulk contribution proportional to the valley Pauli matrix $\tau_1$ is symmetry allowed. Tight-binding calculations yield a bulk coupling more than an order of magnitude larger than typical interface-induced spin-orbit coupling; achieving the same energy scale through the interface-induced mechanism would require electric fields roughly 50 times larger than typical fields. We further derive the symmetry-allowed spin-valley couplings generated by magnetic-field gradients and show how they account for the valley-dependent Zeeman splitting observed in micromagnet experiments. A slight tilt of the background magnetic field out of the plane produces an additional isotropic contribution linear in $B_z$, providing an experimentally accessible signature of the corresponding coupling constant. Finally, we predict a spin-independent micromagnet-induced valley splitting in the $\tau_3$ channel, which is distinct from the $\tau_{1,2}$ channels generated by alloy disorder and therefore remains robust against disorder-induced cancellation. These results establish valley symmetry as a fundamental ingredient in the spin-orbit physics of silicon and provide new mechanisms for controlling and probing spin and valley degrees of freedom in silicon quantum devices.
\end{abstract}



\maketitle
\narrowtext

\begin{figure}[b]
    \centering
    \includegraphics[width=0.85\linewidth]{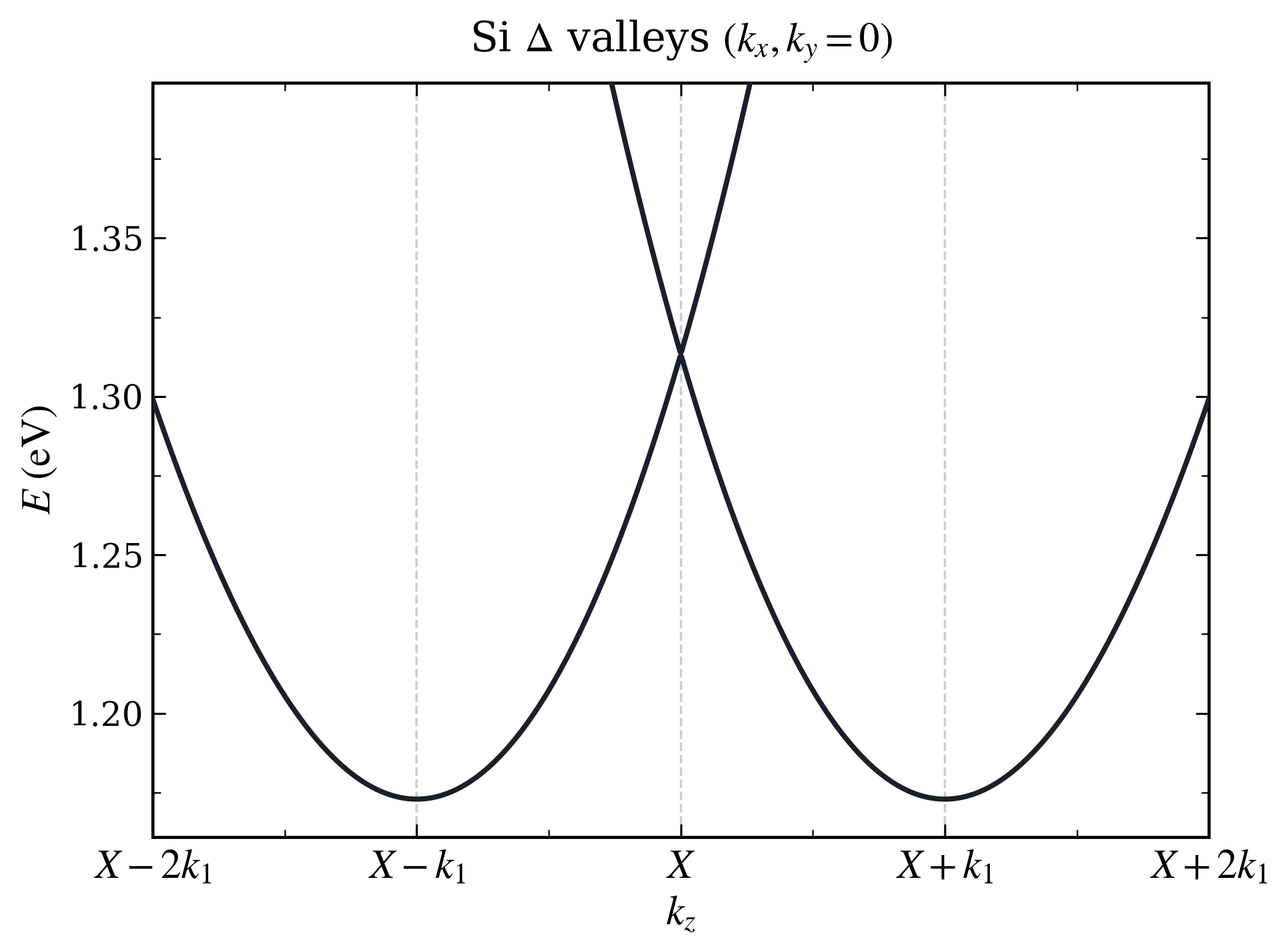}
    \caption{The two valleys in bulk Si along the $k_z$ direction. The X point is $140$ meV above the degenerate conduction band minima.}
    \label{fig:si_valleys}
\end{figure}

\emph{Introduction.} 
Silicon-based heterostructures are a leading platform for spin- and valley-based quantum devices, owing to their long coherence times and compatibility with existing semiconductor technology. In bulk Si, six degenerate conduction-band minima occur on the $\Delta$ line, connecting the center $\Gamma$ of the Brillouin zone to the X point, giving rise to a valley degree of freedom in addition to spin~\cite{YuCardona2010}. In tensile-strained Si/SiGe heterostructures grown along the (001) direction, this bulk degeneracy is partially lifted in the silicon layer, leaving two low-lying valleys $(X_1^+, X_1^-)$ along the growth direction $z$, which split from the other transverse valleys. These valleys, denoted by $|\pm z\rangle$, are located at $\pm \vec{k}_0$, where $\vec{k}_0 = 0.85\vec k_X$. The crossing between the two valleys lies at the X point $\vec{k}_X=\frac{2\pi}{a}(0,0,1)$, with the lattice constant $a \approx 5.43\text{\AA}$ and is protected by the non-symmorphic symmetry of the underlying lattice structure. The residual valley degrees of freedom are offset from the X point by a momentum $\pm \vec k_1\approx \pm 0.15 \vec k_X$ as shown in Fig. \ref{fig:si_valleys}.

One of the central questions in Si-based spin qubits is the nature of the interplay between spin and the valley degree of freedom. Conventional forms of Dresselhaus ($D=k_x\sigma_x-k_y\sigma_y$) and Rashba ($R=k_y\sigma_x-k_x\sigma_y$) SOC rely on extrinsic inversion-symmetry breaking, such as that introduced by sharp interfaces or Ge doping profiles. Both $R$ and $D$ are odd under $z\to -z$ and therefore must be activated by an external inversion-odd quantity, such as the electric field $E_z$. 
The key question addressed in this Letter is whether additional forms of SOC become possible when the valley degree of freedom is treated as an independent quantum degree of freedom with its own transformation properties under symmetry operations.
Our first result is an unexpected \emph{intrinsic bulk} Dresselhaus SOC that does not rely on the presence of interfaces or external electric fields; instead, it is necessarily valley off-diagonal. We extract the magnitude of this SOC using tight-binding calculations. Our analysis is consistent with the local enhancement of the Dresselhaus SOC in Ge-doped systems~\cite{Friesen2023}. Second, we identify an effective coupling between magnetic field gradients and the spin degree of freedom that is exploited for Rabi control via electric fields~\cite{Yoneda2018}. Third, we find that a background micromagnet can also couple directly to the valley degree of freedom without involving spin, providing a new mechanism for valley splitting.

An earlier work~\cite{Prada2011} guided by the assumption that an interface is necessary for $R$ and $D$ SOC where distinct valley wave functions sample the interface differently~\cite{Ferdous2018}, carried out an analysis of the cases with even and odd numbers of layers, leading to symmetry groups that are smaller than the space group of the X point. To reveal the hidden physics of valleys in bulk Si, we go beyond the standard treatment of the valley as a pseudospin index~\cite{Woods2026}. To incorporate well-defined transformation properties for the valley degree of freedom -- encoded in the Pauli matrices $\tau_j$ with $j=1,2,3$~\cite{Jafari2026} -- and its interplay with spin $(\sigma_x,\sigma_y,\sigma_z)$, we fully address the combined \emph{valleyor}$\otimes$\emph{spinor} space as an irreducible representation (IR) of the double space group of the X point. By constructing the double $X$ group and its IRs, we classify all symmetry-allowed terms in the combined spin-valley Hilbert space that can appear in the Hamiltonian, with particular emphasis on SOC terms involving the pair of opposite valleys. Tensile strain in Si/SiGe quantum wells lifts the six-fold valley degeneracy perturbatively while preserving the degeneracy of the two opposite valleys. In analogy with spin, which is a fundamental property of the vacuum and remains a well-defined degree of freedom in solid-state lattices with much smaller symmetry than vacuum, the valley degree of freedom—originating from the $X$ point of the Si Brillouin zone—also remains well defined in strained Si/SiGe quantum wells. 

Our approach establishes that symmetry-allowed bulk Dresselhaus SOC terms necessarily involve a valley flip, $|\pm z\rangle\to|\mp z\rangle$, whereas Rashba SOC couples identically to both valleys. While the latter retains the same form in electronic systems without a valley degree of freedom, the former emerges only when the valley degree of freedom is taken into account. Thus, the fundamental symmetry of the $X$ point enables a form of bulk SOC that has no counterpart in valley-less systems. 
This fundamental symmetry also provides a natural framework for addressing inhomogeneous systems, offering an explanation for recent simulations involving Ge~\cite{Friesen2023}, which indicate that the double $X$ symmetry is highly robust and can govern local fields as well.
The significance of our findings lies in establishing that the presence of the valley degree of freedom fundamentally modifies the ways in which different degrees of freedom can be coupled. Specifically, we (1) identify an \emph{intrinsic} mechanism for spin-orbit-valley (SOV) coupling in \emph{bulk} Si that goes beyond interface-induced intervalley scattering~\cite{Nestoklon2008,Saraiva2009,Ruskov2018}; (2) explain how magnetic field gradients enable Rabi control of the spin state~\cite{Yoneda2018}; and (3) demonstrate that these gradients additionally contribute to the generation of valley splitting.


\emph{The space group at the X point and its double group.}
The point group for the Si structure (based on the tetrahedral group) is given by $D_{2d}=\{E,C^2_{4z}, C^2_{4x}, C^2_{4y}, 2S_{4z},m_1,m_2\}$ which is defined by its action on the coordinates $xyz$ as $\{xyz, \bar{x}\bar{y}z, x\bar{y}\bar{z}, \bar{x}y\bar{z}, \bar{y}x\bar{z}, y\bar{x}\bar{z}, yxz, \bar{y}\bar{x}z\}$~\cite{Jafari2026}.  
The non-symmorphic space group is then obtained by introducing a glide operator $g = Tm_z$ which consists of a mirror $m_z: xyz\rightarrow xy\bar{z}$ and a  fractional translation $T = \frac{a}{4}(1,1,1)$. To form a closed group, one also needs to include the pure translation $Q = \frac{a}{2}(0, -1, -1)$  as well as the product $Qg$, resulting in a 32-element space group. This closes when acting on Bloch waves $\psi = e^{i\vec{k}_X\vec{r}}u(\vec{r}) = e^{i2\pi z/a}u(\vec{r})$~\cite{YuCardona2010}.
The double group contains all of the symmetry operations of the single group, with an added rotation $\mathcal{R}$ by $2\pi$, which only affects the spinor part $(\chi^+, \chi^-)$ of the Bloch wave. Compared to the single group, which has 14 conjugacy classes, the double group contains five additional classes and hence five new IRs, which means that their dimensions should satisfy the criterion $\sum_{i=15}^{19} l_i^2 = 32$. This constraint has three solutions: $(2,2,2,2,4)$, $(1,1,1,2,5)$, and $(1,2,3,3,3)$. Only the first set leads to a consistent character table. One of the two-dimensional representations, which we denote by $Y_1$, naturally describes the pure spinor representation, whereas the four-dimensional representation ($Y_{5}$) describes the spinor$\otimes$\textit{valleyor} space. This is consistent only with the solution $(2,2,2,2,4)$ for the description of the spin-valley coupling in Si-based spin qubits.


We can calculate the transformation properties of various objects such as components of the strain tensor, polar vectors (e.g. wave vector $\vec{k}$, or electric field $\vec E$), a pseudovector $\vec{R}$ (e.g. magnetic field $\vec B$), and the valley and spin Pauli matrices and assign them to their respective IRs. For more details, see supplementary information (SI). The $x$ and $y$ components of (pseudo-)vectors belong to two-dimensional IRs: The doublet $(x,y)$ belongs to the $M_5$ IR and $(\sigma_x, \sigma_y)$ belong to the two-dimensional $M'_5$ IR, while $\sigma_z$ belongs to $M_3$. This behavior contrasts with that of the valley Pauli matrices $\tau_j$, which all transform according to \emph{three distinct one-dimensional} IRs~\cite{Jafari2026}. 
It is important to note that, unlike the three spin Pauli matrices, the three valley Pauli matrices are not associated with the spatial directions $x,y,z$. For this reason, we label the valley Pauli matrices with numbers $1,2,3$. This contrast reinforces the view that the two-component object representing the relative weights and phases of the $|\pm z\rangle$ valleys, which we dub the valleyor, is fundamentally distinct from a spinor. In fact, spinors transform according to the IR $Y_1$, whereas the two-component valleyor transforms according to the IR $X_1$; the individual valley Pauli matrices $\tau_1$, $\tau_2$, and $\tau_3$ transform according to three distinct one-dimensional IRs. We refer the reader to the SI for the multiplication table of the IRs of the double group. 

Symmetry-allowed terms are those combinations of IR products that transform according to the scalar IR $M_1$, thereby ensuring the invariance of the Hamiltonian under the full double $X$ symmetry. 
Permissible terms in the Hamiltonian must also be even under time-reversal (TR) symmetry. In the space of valley states $|\pm z\rangle$, the TR operator is given by $T = \hat{\tau}_1 K$, where $K$ denotes complex conjugation, leading to~\cite{Jafari2026}
\begin{equation}
   T\tau_1T^{-1} = +\tau_1,~~ T\tau_2T^{-1} = +\tau_2,~~ T\tau_3T^{-1} = -\tau_3. 
\end{equation}
Note that for the spin $\sigma_i$, \emph{all} components are odd under TR. Among other physically relevant quantities, $\vec{k}$ and $\vec{B}$ are also TR-odd, whereas $\vec{E}$ and strain $\varepsilon$ are TR-even. These symmetry considerations enable us to systematically construct all Hamiltonians that are invariant under the double X group.
\begin{table}[t]
    \centering
    \begin{tabular}{l | l}
         \hline
        Spin-orbit couplings & Type\\
        \hline
        $\alpha_0 E_z ( k_x \sigma_y - k_y \sigma_x )\tau_0$ & Rashba, Interface \\
        $\beta_2E_z ( k_x \sigma_x - k_y \sigma_y ) \tau_2$ & Dresselhaus, Interface \\
        $\beta_1( k_x \sigma_x - k_y \sigma_y ) \tau_1$ & Dresselhaus, Bulk 
    \end{tabular}
    \caption{Permissible spin-orbit terms. $\tau_0$ is the unit $2\times 2$ matrix and $\tau_j$ with $j=1,2,3$ are Pauli matrices in valley space.
    The subscript of the coefficients $\alpha,\beta$ is chosen to match that of the valley matrix.}
    \label{tab:spin_orbit_all}
\end{table}

\emph{Nature of spin-orbit in the presence of valleys.}
Up to first order in external fields for an unstrained system with TR symmetry, three distinct SOC terms arise (Tab.~\ref{tab:spin_orbit_all}). The Rashba SOC is proportional to the identity matrix $\tau_0$ in valley space, meaning that it has the same form for both valleys, whereas the Dresselhaus SOC necessarily involves off-diagonal valley Pauli matrices; in particular, symmetry forbids any Dresselhaus term proportional to $\tau_0$. The first two terms in Tab.~\ref{tab:spin_orbit_all} are extrinsic interface-induced contributions that require $z \to -z$ symmetry breaking enforced via an electric field $E_z$, while the last term is a bulk contribution. 
The distinction between the two types of Dresselhaus SOC arises from the different irreducible representations, or symmetry types, of $\tau_1$ and $\tau_2$. 
In particular, the combination $\tau_2 E_z$ transforms in the same way as $\tau_1$. 
Conversely, the distinct symmetry of $\tau_1$ permits a pure $\tau_1$ Dresselhaus term even in the absence of $E_z$, so that a Dresselhaus-type SOC term can be constructed without requiring any explicit interface contribution or external field. This highlights the crucial role of the transformation properties of the valley Pauli matrices in generating SOC terms that have no counterpart in valley-less systems.

In order to estimate the strength of the extrinsic SOC constants $\alpha_0$ and $\beta_2$ in the first two rows of Tab.~\ref{tab:spin_orbit_all} and establish the necessity of $\tau_2 E_z$ combination for the Dresselhaus term, we rely on Ref.~\cite{Friesen2023} where the authors analyze the Hamiltonian $H_{\rm SO} = \alpha R + \beta D$ when both $D$ and $R$ terms are induced by $E_z$. Their Hamiltonian can be obtained from a minimal double group based on the point group $C_{2v}$. Two geometries are considered: a Si quantum well confined between Si$_{0.7}$Ge$_{0.3}$ barriers, and the same structure but incorporating Ge concentration oscillations of wavelength $\lambda$ (the “wiggle well”). They employed tight-binding simulations to compute the Rashba and Dresselhaus coefficients associated with the ground and first excited states. For the quantum well without Ge concentration oscillations, both Rashba and Dresselhaus interactions are found to depend linearly on the vertical electric field $E_z$~\cite{Friesen2023} (cf. rows $1,2$ in Tab.~\ref{tab:spin_orbit_all}). The Rashba coefficients for the two valleys have the same sign (cf. Fig. 3b~\cite{Friesen2023}), confirming that the $R$ term must be proportional to $\tau_0$, whereas the Dresselhaus coefficients exhibit opposite signs for the ground and excited states (cf. Fig. 3a~\cite{Friesen2023}), implying that they must have an $E_z\tau_j$ form. Our analysis based on the double X group shows that only the $j=2$ Pauli matrix gives an invariant SOC as in the second row of Tab.~\ref{tab:spin_orbit_all}. In our framework, $E_z D$ belongs to the $M_4$ IR and must therefore be multiplied by a $M_4$ valley Pauli matrix (i.e. $\tau_2$) to produce a permissible scalar.

\begin{figure}[t]
    \centering
    \subfloat[]{
        \includegraphics[width=0.47\linewidth]{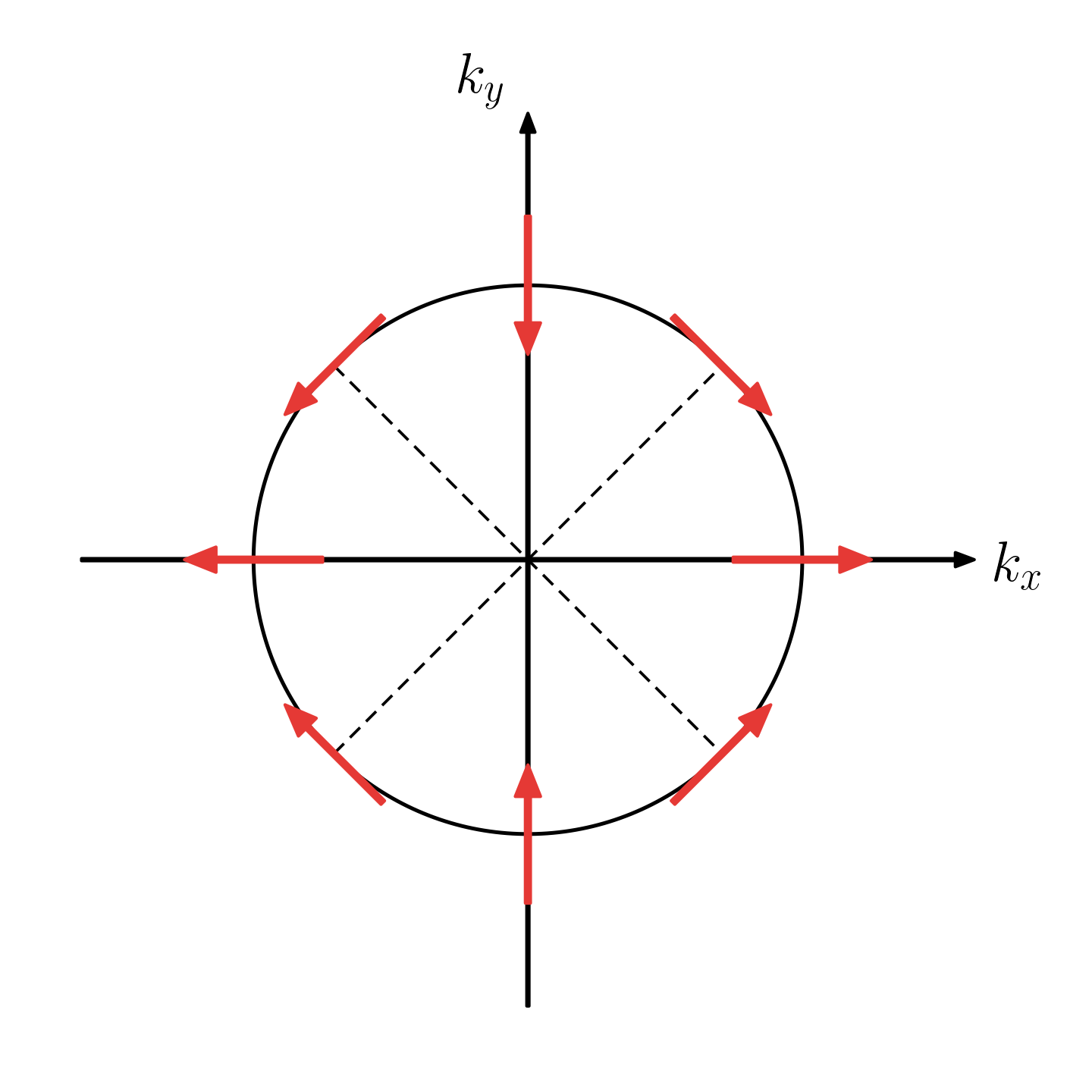}
        \label{fig:dresselhaus}
    }\hfill
    \subfloat[]{
        \includegraphics[width=0.47\linewidth]{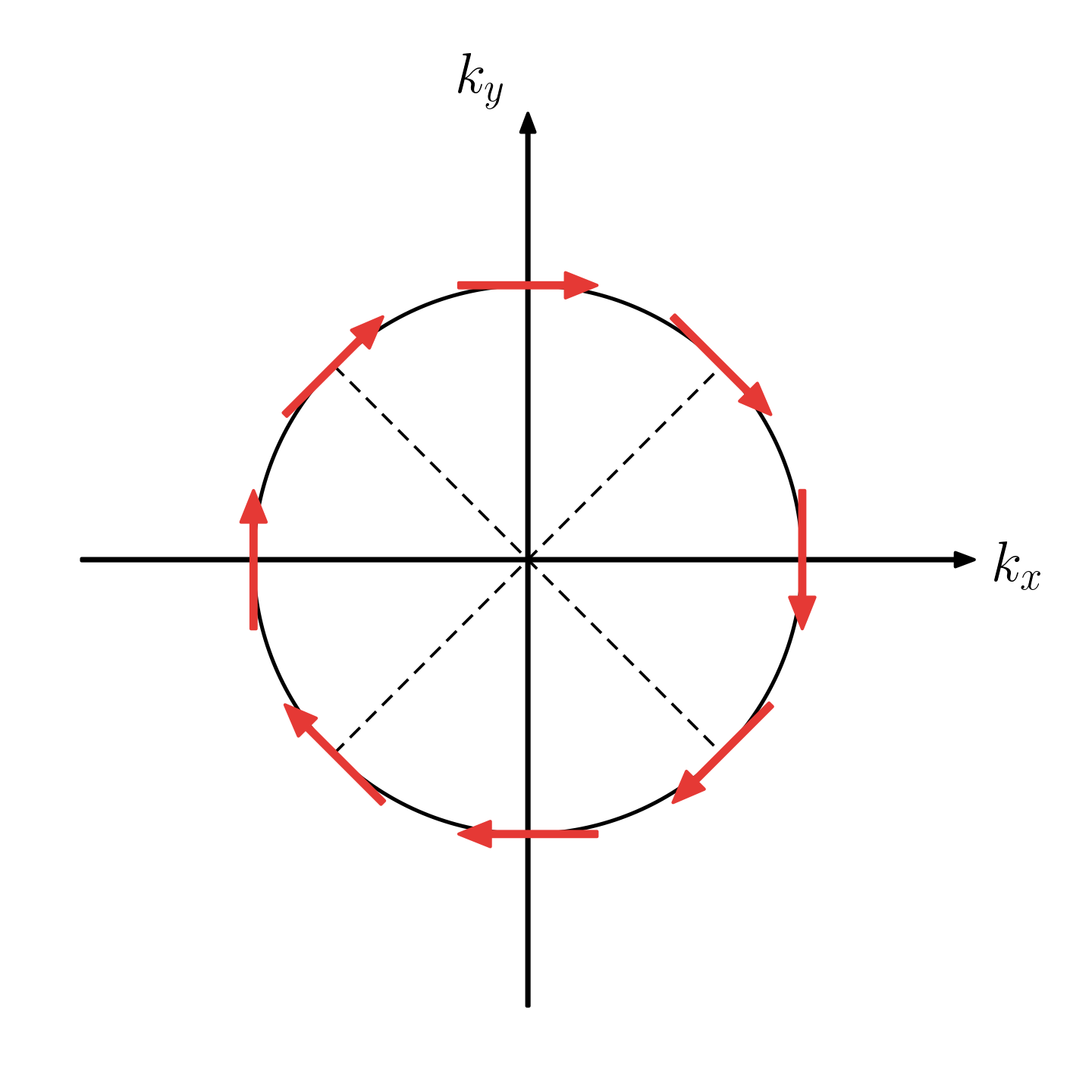}
        \label{fig:rashba}
    }
    \caption{Spin textures on the Fermi surface near one of the valleys for (a) Dresselhaus and (b) Rashba spin-orbit coupling. For Dresselhaus coupling, the spin texture in the opposite valley has the reversed orientation, whereas the Rashba spin texture is identical in both valleys.
}
    \label{fig:spin_textures}
\end{figure}

From the data in Fig. 3 of~\cite{Friesen2023}, the coefficient $\beta_2$ associated with the Dresselhaus term in the second row of Tab.~\ref{tab:spin_orbit_all} is $\beta_2 \approx 2 \mu\text{eV nm}/\text{(mV/nm)}$, and the coefficient of the Rashba term in the first row is $\sim$ 7 times smaller. The spin textures corresponding to Rashba and Dresselhaus SOC for $\tau_a$-eigenstates ($a=0,2$) at energies (Fermi surfaces) above the bottom of the conduction bands are shown in Fig.~\ref{fig:spin_textures}. Beyond their potential relevance to transport in bulk Si, these textures can also influence spin qubits in quantum-dot geometries. In particular, zero-point motion allows the momentum-dependent terms to generate effective spin-valley coupling when the momentum matrix elements are integrated out~\cite{Woods2026}.

Our framework can also explain a very small asymmetry in the Dresselhaus coefficient observed in Fig.~3a of Ref.~\cite{Friesen2023}, where the sign between the two valleys is not perfectly opposite. This suggests the presence of a tiny Dresselhaus component proportional to $\tau_0$, a term that is not permitted under the full symmetry of Si. Such a term can be enabled if the symmetry is reduced by eliminating the operations ${ C_{4x}^2, C_{4y}^2, S_{4z}, S_{4z}^{-1} }$. This reduction of the underlying point group to $C_{2v}$ permits a term of the form $E_z D \tau_0$~\cite{Prada2011}. Such partial symmetry breaking can be attributed to the geometry of the quantum well or to the Ge doping profile. This indicates that, despite symmetry lowering by the Ge doping profile, the double X symmetry remains dominant, and deviations from this symmetry give rise to sub-leading corrections. In the extreme limit of a wiggle-well configuration, where the Ge concentration wavelength is $\lambda = a/2 \approx 0.27\ \text{nm}$, the Dresselhaus coupling exhibits a sharp peak, acquires the same sign for both valleys, and becomes independent of $E_z$~\cite{Friesen2023}. This behavior can also be understood within our group-theoretical framework: a modulation period of $\lambda = a/2$ implies that only one of the fcc sublattices is doped with Ge, thereby rendering the two sublattices inequivalent and reducing the space-group symmetry to the point group $D_{2d}$ that dominates the symmetry of the medium. Within this symmetry descent, the term $D$ alone becomes a scalar and therefore need not couple to an electric field or to a valley Pauli matrix.

The valley-flip character of Dresselhaus SOC becomes more evident and natural at the atomic scale in the presence of Ge doping. An isolated Ge atom, acting as a localized scatterer, inherently induces a valley flip, thereby suggesting an X symmetry-compatible local SOV coupling for the wiggle-well potentials $\{V_1,V_2\}=\alpha_{{1,2}}^{\rm Ge}(\lambda)(x\sigma_x - y\sigma_y)\{\tau_1,\tau_2 E_z\}$, where $\lambda$ denotes the wavelength of the Ge concentration~\cite{Friesen2023}. Such a local SOV coupling is expected to remain valid for generic incommensurate $\lambda \ne a/2$, for which X symmetry continues to be the governing symmetry. 

\begin{figure}[t]
    \centering
    \includegraphics[width=0.75\linewidth]{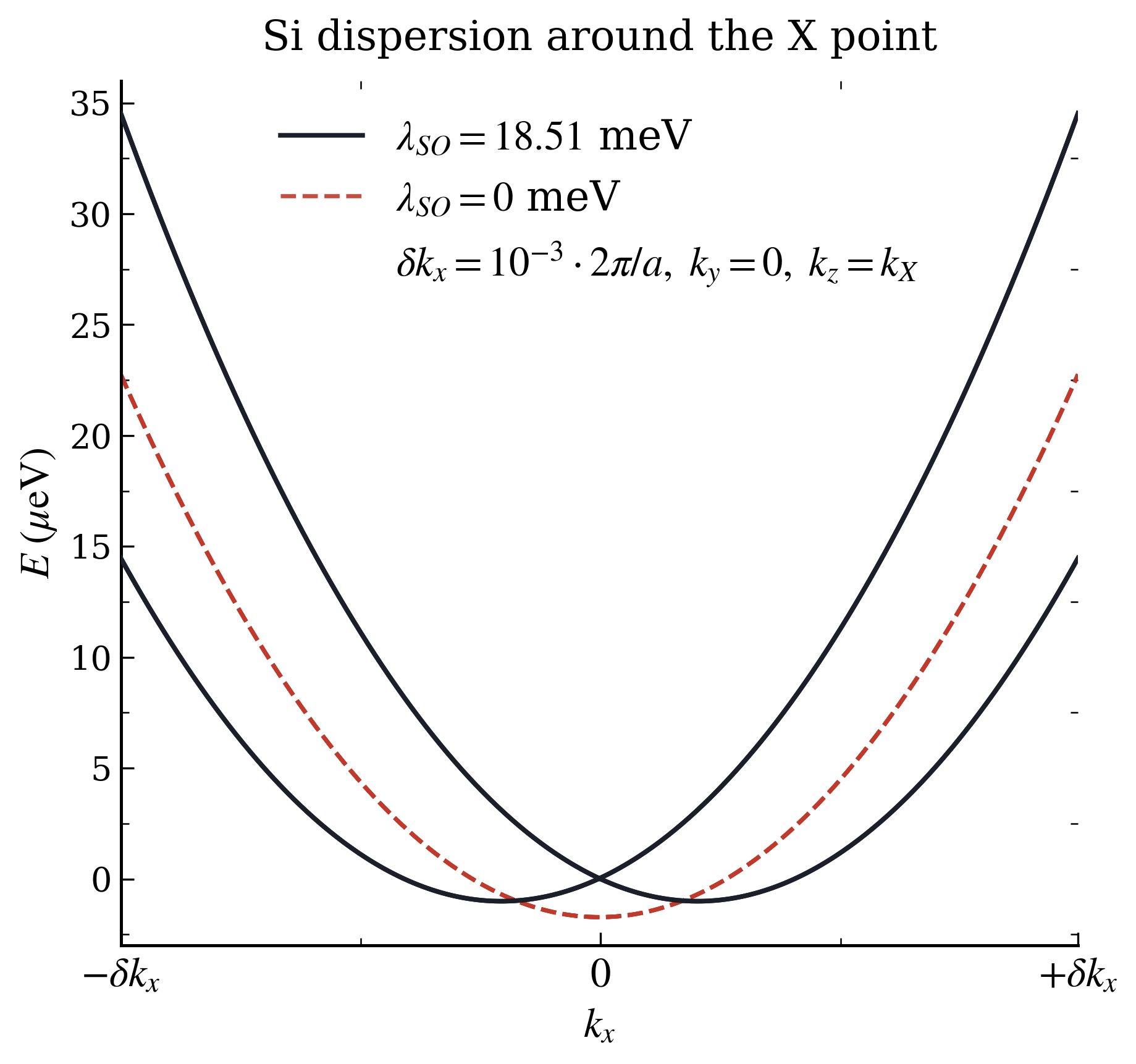}
    \caption{Tight-binding dispersion along the $k_x$ direction around the $X$ point, shown for $\lambda_{\rm SO} = 18.51$ meV (black) and $\lambda_{\rm SO} = 0$ (red), using the parameters of~\cite{Niquet2009}. The range of $k_x$ is $\pm\delta k_x = \pm 10^{-3}\cdot 2\pi/a$. The energy axis is offset by $1.3133$ eV for clarity. Note that, in the tight-binding calculation, the $X$ point corresponds to $k_z=2\pi/a$, whereas in our matrix $k.p$ theory, which is expanded around the $X$ point itself, it is represented by $k_z=0$.
}
    \label{fig:x_point_SO}
\end{figure}

\emph{Bulk Dresselhaus-valley term.}
 To determine the coefficient $\beta_1$ of the bulk term $\beta_1D\tau_1$, following \cite{Niquet2009}, we implement a tight-binding simulation of Si quantum wells. We use an $sp^3d^5s^*$ model with onsite spin-orbit interaction between $p$ orbitals, controlled by the parameter $\lambda_{\rm SO} \approx 18.5$ meV~\cite{Niquet2009}.
In the conduction bands near the X point, the dominant term in the Hamiltonian is $-\hbar k_z v_1\tau_3$ where $v_1=(\hbar k_1/m_\parallel)$ is a velocity scale associated with the offset $k_1$ in Fig.~\ref{fig:si_valleys}. The momenta are measured from the X point, such that exactly at the X point, there is a four-fold degeneracy. The spin-orbit parameter $\lambda_{\rm SO}$ lifts the spin degeneracy at nonzero momenta, causing a crossing in the $k_x$ and $k_y$ directions induced by lateral shift of the bands in Fig.~\ref{fig:x_point_SO}.  
Using $m_\parallel \approx 0.98m_e$, the slope of this crossing at the \emph{X point} is $\beta_1^{\rm X}\approx 0.87$ meV nm. 

\begin{table}[t]
    \centering
    \begin{tabular}{l|l}
        \hline
        Term & index \\ \hline
        $\lambda_1( \partial_x B_z \sigma_y + \partial_y B_z \sigma_x )\tau_1$ & $1$ \\        $\lambda_2(\partial_xB_y+\partial_yB_x)\sigma_z\tau_1$ & $2$ \\
        $\lambda_3(\partial_zB_x\sigma_y + \partial_zB_y\sigma_x) \tau_1$  & $3$\\
    \end{tabular}
    \caption{Symmetry-allowed spin-valley couplings in an inhomogeneous micromagnet field. Term $3$ has negligible derivatives in the experiment of Ref.~\cite{Ferdous2018}.}
    \label{tab:B-field_terms}
\end{table}

In order to estimate the magnitude of the bulk $\beta_1$ at the valleys (conduction band minima) $k_z = \pm k_1$, it is useful to numerically compute the difference of the eigenvalues that is caused by SOC, $H(\lambda_{\rm SO}\approx 18.5\text{ meV}) - H(\lambda_{\rm SO}=0)$ that eliminates the effects of any $\tau_0$-dependent terms in the Hamiltonian. These are both shown \textit{separately} in Fig.~\ref{fig:x_point_SO}. Diagonalizing $-\hbar k_z v_1\tau_3 + \beta_1( k_x \sigma_x - k_y \sigma_y ) \tau_1$ and subtracting the $\lambda_{\rm SO}=0$ contribution, we obtain the following expression for the \textit{difference} between the red and black curves:
\begin{equation*}
    E(\vec{k}) = E_0 \pm \left(\sqrt{\beta_1^2( k_x \sigma_x - k_y \sigma_y )^2 + \left(\hbar k_z v_1\right)^2} - \hbar k_z v_1\right),
\end{equation*}
where a constant $E_0$ accounts for any higher-order energy shift.
This function with free parameters $E_0$ and $\beta_1$ can be fitted to the tight-binding results for a range of values of $k_z$. 
Around the $X$ point, which in our $k.p$ theory is specified by $|k_x|<\delta k_x$, $k_y=0$, and $k_z=0$, the above equation yields a linear dependence on $k_x$ for the difference between the red and black curves. The difference between the black and red quadratic dispersions in Fig.~\ref{fig:x_point_SO} is also well described by this linear dependence, whose slope gives $\beta_1^{\rm X}\approx 0.87$ meV nm. 
The same fitting procedure can be applied at arbitrary $\vec k$ values, including those near the conduction-band minima. The resulting values of $\beta_1$ vary slightly from those obtained at the $X$ point; at the bottom of the valley, we find $\beta_1^{\pm k_1}\approx0.67$ meV nm. 
Achieving this energy scale through interface-induced Dresselhaus SOC~\cite{Friesen2023} would require an electric field strength on the order of $0.5$ V/nm, approximately $50$ times larger than the typical fields of $\sim 10$ mV/nm.
Spin-orbit coupling in silicon is known to be weak, but has been predicted theoretically~\cite{Guo2005,Tetlow2013} and found experimentally~\cite{Ezhevskii2020} to yield a small but non-negligible spin Hall angle of $\theta_{\rm SH} \sim 1\times 10^{-4}$. This is likely to be caused by the bulk coupling $\beta_1(\vec k)$.

\emph{Coupling to a micromagnet.} 
Motivated by experiments on Rabi control in non-uniform magnetic fields~\cite{Yoneda2018}, we identify three distinct terms that couple the spin $\sigma_k$ to magnetic-field gradients $\partial_i B_j$, in the form $\partial_i B_j\sigma_k\tau_l$, as summarized in Tab.~\ref{tab:B-field_terms}. Remarkably, all three terms are necessarily accompanied by the valley-flip operator $\tau_1$.
Such inhomogeneous fields are realized experimentally using micromagnets~\cite{Kawakami2014,Yoneda2018,Ferdous2018}. 
In~\cite{Ferdous2018}, the effects of $\partial_z B_i$ are negligible, so the third row in Tab.~\ref{tab:B-field_terms} can be ignored.
There is an additional constant external in-plane magnetic field $\vec{B}_{\rm ext}$ in the background, which is stronger than the micromagnets' field by a factor $\sim\!5$, weakly depending on its in-plane angle $\theta$. Using this set-up, a difference $\Delta_\tau E_Z$ in the Zeeman splitting between the valley eigenstates is observed~\cite{Ferdous2018}.

A dominant Zeeman coupling yields the same spin splitting $E_Z$ for both valleys. However, the gradient terms introduce valley-dependent corrections to $E_Z$, which for the two couplings in Tab.~\ref{tab:B-field_terms} give the \textit{difference} between the Zeeman splitting of the two valleys $\Delta_\tau E_Z^1 \approx 4\lambda_1(B_x\partial_yB_z+B_y\partial_xB_z)/|\vec{B}|$ 
and  $\Delta_\tau E_Z^2 \approx 4\lambda_2B_z(\partial_xB_y+\partial_yB_x)/|\vec{B}|$. 
In Fig.~4d of Ref.~\cite{Ferdous2018}, the contribution of the inhomogeneous $B$-field only very weakly depends on $|\vec{B}_{\text{ext}}|$. Approximating $|\vec{B}| \approx |\vec{B}_{\text{ext}}|$, this 
can be explained by the above $\Delta_\tau E_Z^1$ term when the magnetic field is in-plane, e.g. $B_x=|\vec B|\cos\phi$ from which $|\vec B|$ drops out.
Using Fig.~1c of Ref.~\cite{Ferdous2018} and data for the magnitude of the inhomogeneous $B$-field, we estimate $\lambda_1\approx 13\mu $eV nm/T.
The $\Delta_\tau E_Z^2$ term predicts that when a dominantly in-plane magnetic field involves a deliberate slight tilt to produce a $B_z$ component, an additional contribution linear in $B_z$ and independent of the direction of the in-plane component arises. These two properties allow us to experimentally isolate this term and measure the coupling $\lambda_2$. 

The effects of a micromagnet are not limited to valley-dependent spin splitting. One can also consider terms of the form $\partial_i B_j \sigma_0 \tau_l$~\cite{Jafari2026}, which act independently of the spin degree of freedom.
This analysis gives a micromagnet-induced valley-splitting term of the form ${\cal B}_3\tau_3=(\partial_x B_y - \partial_y B_x)\tau_3$. 
Being proportional to $\tau_3$ protects the micromagnet contribution to valley-splitting from cancellation by disorder-induced valley splitting and prevents it from being compensated by SiGe alloy disorder, as in Ref.~\cite{Losert2023}. Since alloy disorder contributes only to the $\tau_1$ and $\tau_2$ components, our new $\tau_3$ term provides a lower bound on the valley splitting in the presence of a magnetic-field gradient.


\emph{Conclusions.}
In the band limit of Si, Dresselhaus SOC is always accompanied by valley-off-diagonal matrices, i.e., it is inherently a SOV coupling. 
The $\tau_2$ channel arises from inversion-breaking effects such as an interface, whereas the $\tau_1$ channel exists intrinsically in the bulk. 
In contrast, Rashba SOC is strictly interface-induced and remains valley-diagonal, i.e., proportional to $\tau_0$. 
Our tight-binding calculation shows that the bulk Dresselhaus coupling is more than an order of magnitude larger than 
the typical interface-induced SOC in spin qubit quantum dots; achieving the same energy scale through the interface-induced mechanism would require an electric field approximately 50 times larger than typical fields. Our framework (1) explains the valley-dependent spin splitting 
in the presence of a micromagnet, where valley states are eigenstates of $\tau_1$; 
(2) predicts an isotropic $B_z$-linear extension of this effect when the background magnetic field acquires an out-of-plane component; and 
(3) predicts that micromagnets can also generate valley splitting in the $\tau_3$ channel, which is distinct from the $\tau_1$ and $\tau_2$ channels generated by alloy disorder and therefore remains robust against disorder-induced cancellation.

\emph{Acknowledgments}. 
We have used ChatGPT (OpenAI) to rephrase and improve the clarity and flow of the manuscript.

\bibliography{references}

\begin{thebibliography}{16}%
\makeatletter
\providecommand \@ifxundefined [1]{%
 \@ifx{#1\undefined}
}%
\providecommand \@ifnum [1]{%
 \ifnum #1\expandafter \@firstoftwo
 \else \expandafter \@secondoftwo
 \fi
}%
\providecommand \@ifx [1]{%
 \ifx #1\expandafter \@firstoftwo
 \else \expandafter \@secondoftwo
 \fi
}%
\providecommand \natexlab [1]{#1}%
\providecommand \enquote  [1]{``#1''}%
\providecommand \bibnamefont  [1]{#1}%
\providecommand \bibfnamefont [1]{#1}%
\providecommand \citenamefont [1]{#1}%
\providecommand \href@noop [0]{\@secondoftwo}%
\providecommand \href [0]{\begingroup \@sanitize@url \@href}%
\providecommand \@href[1]{\@@startlink{#1}\@@href}%
\providecommand \@@href[1]{\endgroup#1\@@endlink}%
\providecommand \@sanitize@url [0]{\catcode `\\12\catcode `\$12\catcode
  `\&12\catcode `\#12\catcode `\^12\catcode `\_12\catcode `\%12\relax}%
\providecommand \@@startlink[1]{}%
\providecommand \@@endlink[0]{}%
\providecommand \url  [0]{\begingroup\@sanitize@url \@url }%
\providecommand \@url [1]{\endgroup\@href {#1}{\urlprefix }}%
\providecommand \urlprefix  [0]{URL }%
\providecommand \Eprint [0]{\href }%
\providecommand \doibase [0]{https://doi.org/}%
\providecommand \selectlanguage [0]{\@gobble}%
\providecommand \bibinfo  [0]{\@secondoftwo}%
\providecommand \bibfield  [0]{\@secondoftwo}%
\providecommand \translation [1]{[#1]}%
\providecommand \BibitemOpen [0]{}%
\providecommand \bibitemStop [0]{}%
\providecommand \bibitemNoStop [0]{.\EOS\space}%
\providecommand \EOS [0]{\spacefactor3000\relax}%
\providecommand \BibitemShut  [1]{\csname bibitem#1\endcsname}%
\let\auto@bib@innerbib\@empty
\bibitem [{\citenamefont {Yu}\ and\ \citenamefont
  {Cardona}(2010)}]{YuCardona2010}%
  \BibitemOpen
  \bibfield  {author} {\bibinfo {author} {\bibfnamefont {P.~Y.}\ \bibnamefont
  {Yu}}\ and\ \bibinfo {author} {\bibfnamefont {M.}~\bibnamefont {Cardona}},\
  }\href {https://doi.org/10.1007/978-3-642-00710-1} {\emph {\bibinfo {title}
  {Fundamentals of Semiconductors: Physics and Materials Properties}}},\
  \bibinfo {edition} {4th}\ ed.,\ Graduate Texts in Physics\ (\bibinfo
  {publisher} {Springer},\ \bibinfo {address} {Berlin, Heidelberg},\ \bibinfo
  {year} {2010})\BibitemShut {NoStop}%
\bibitem [{\citenamefont {Woods}\ \emph {et~al.}(2023)\citenamefont {Woods},
  \citenamefont {Eriksson}, \citenamefont {Joynt},\ and\ \citenamefont
  {Friesen}}]{Friesen2023}%
  \BibitemOpen
  \bibfield  {author} {\bibinfo {author} {\bibfnamefont {B.~D.}\ \bibnamefont
  {Woods}}, \bibinfo {author} {\bibfnamefont {M.~A.}\ \bibnamefont {Eriksson}},
  \bibinfo {author} {\bibfnamefont {R.}~\bibnamefont {Joynt}},\ and\ \bibinfo
  {author} {\bibfnamefont {M.}~\bibnamefont {Friesen}},\ }\href
  {https://doi.org/10.1103/PhysRevB.107.035418} {\bibfield  {journal} {\bibinfo
   {journal} {Physical Review B}\ }\textbf {\bibinfo {volume} {107}},\ \bibinfo
  {pages} {035418} (\bibinfo {year} {2023})}\BibitemShut {NoStop}%
\bibitem [{\citenamefont {Yoneda}\ \emph {et~al.}(2018)\citenamefont {Yoneda},
  \citenamefont {Takeda}, \citenamefont {Otsuka}, \citenamefont {Nakajima},
  \citenamefont {Delbecq}, \citenamefont {Allison}, \citenamefont {Honda},
  \citenamefont {Kodera}, \citenamefont {Oda}, \citenamefont {Hoshi},
  \citenamefont {Usami}, \citenamefont {Itoh},\ and\ \citenamefont
  {Tarucha}}]{Yoneda2018}%
  \BibitemOpen
  \bibfield  {author} {\bibinfo {author} {\bibfnamefont {J.}~\bibnamefont
  {Yoneda}}, \bibinfo {author} {\bibfnamefont {K.}~\bibnamefont {Takeda}},
  \bibinfo {author} {\bibfnamefont {T.}~\bibnamefont {Otsuka}}, \bibinfo
  {author} {\bibfnamefont {T.}~\bibnamefont {Nakajima}}, \bibinfo {author}
  {\bibfnamefont {M.~R.}\ \bibnamefont {Delbecq}}, \bibinfo {author}
  {\bibfnamefont {G.}~\bibnamefont {Allison}}, \bibinfo {author} {\bibfnamefont
  {T.}~\bibnamefont {Honda}}, \bibinfo {author} {\bibfnamefont
  {T.}~\bibnamefont {Kodera}}, \bibinfo {author} {\bibfnamefont
  {S.}~\bibnamefont {Oda}}, \bibinfo {author} {\bibfnamefont {Y.}~\bibnamefont
  {Hoshi}}, \bibinfo {author} {\bibfnamefont {N.}~\bibnamefont {Usami}},
  \bibinfo {author} {\bibfnamefont {K.~M.}\ \bibnamefont {Itoh}},\ and\
  \bibinfo {author} {\bibfnamefont {S.}~\bibnamefont {Tarucha}},\ }\href
  {https://doi.org/10.1038/s41565-017-0014-x} {\bibfield  {journal} {\bibinfo
  {journal} {Nature Nanotechnology}\ }\textbf {\bibinfo {volume} {13}},\
  \bibinfo {pages} {102} (\bibinfo {year} {2018})}\BibitemShut {NoStop}%
\bibitem [{\citenamefont {Prada}\ \emph {et~al.}(2011)\citenamefont {Prada},
  \citenamefont {Klimeck},\ and\ \citenamefont {Joynt}}]{Prada2011}%
  \BibitemOpen
  \bibfield  {author} {\bibinfo {author} {\bibfnamefont {M.}~\bibnamefont
  {Prada}}, \bibinfo {author} {\bibfnamefont {G.}~\bibnamefont {Klimeck}},\
  and\ \bibinfo {author} {\bibfnamefont {R.}~\bibnamefont {Joynt}},\ }\href
  {https://doi.org/10.1088/1367-2630/13/1/013009} {\bibfield  {journal}
  {\bibinfo  {journal} {New Journal of Physics}\ }\textbf {\bibinfo {volume}
  {13}},\ \bibinfo {pages} {013009} (\bibinfo {year} {2011})}\BibitemShut
  {NoStop}%
\bibitem [{\citenamefont {Ferdous}\ \emph {et~al.}(2018)\citenamefont
  {Ferdous}, \citenamefont {Kawakami}, \citenamefont {Scarlino}, \citenamefont
  {Nowak}, \citenamefont {Ward}, \citenamefont {Savage}, \citenamefont
  {Lagally}, \citenamefont {Coppersmith}, \citenamefont {Friesen},
  \citenamefont {Eriksson}, \citenamefont {Vandersypen},\ and\ \citenamefont
  {Rahman}}]{Ferdous2018}%
  \BibitemOpen
  \bibfield  {author} {\bibinfo {author} {\bibfnamefont {R.}~\bibnamefont
  {Ferdous}}, \bibinfo {author} {\bibfnamefont {E.}~\bibnamefont {Kawakami}},
  \bibinfo {author} {\bibfnamefont {P.}~\bibnamefont {Scarlino}}, \bibinfo
  {author} {\bibfnamefont {M.~P.}\ \bibnamefont {Nowak}}, \bibinfo {author}
  {\bibfnamefont {D.~R.}\ \bibnamefont {Ward}}, \bibinfo {author}
  {\bibfnamefont {D.~E.}\ \bibnamefont {Savage}}, \bibinfo {author}
  {\bibfnamefont {M.~G.}\ \bibnamefont {Lagally}}, \bibinfo {author}
  {\bibfnamefont {S.~N.}\ \bibnamefont {Coppersmith}}, \bibinfo {author}
  {\bibfnamefont {M.}~\bibnamefont {Friesen}}, \bibinfo {author} {\bibfnamefont
  {M.~A.}\ \bibnamefont {Eriksson}}, \bibinfo {author} {\bibfnamefont
  {L.~M.~K.}\ \bibnamefont {Vandersypen}},\ and\ \bibinfo {author}
  {\bibfnamefont {R.}~\bibnamefont {Rahman}},\ }\bibfield  {journal} {\bibinfo
  {journal} {npj Quantum Information}\ }\textbf {\bibinfo {volume} {4}},\ \href
  {https://doi.org/10.1038/s41534-018-0075-1} {10.1038/s41534-018-0075-1}
  (\bibinfo {year} {2018})\BibitemShut {NoStop}%
\bibitem [{\citenamefont {Woods}\ \emph {et~al.}(2026)\citenamefont {Woods},
  \citenamefont {Losert}, \citenamefont {Joynt},\ and\ \citenamefont
  {Friesen}}]{Woods2026}%
  \BibitemOpen
  \bibfield  {author} {\bibinfo {author} {\bibfnamefont {B.~D.}\ \bibnamefont
  {Woods}}, \bibinfo {author} {\bibfnamefont {M.~P.}\ \bibnamefont {Losert}},
  \bibinfo {author} {\bibfnamefont {R.}~\bibnamefont {Joynt}},\ and\ \bibinfo
  {author} {\bibfnamefont {M.}~\bibnamefont {Friesen}},\ }\bibfield  {journal}
  {\bibinfo  {journal} {Physical Review Letters}\ }\textbf {\bibinfo {volume}
  {136}},\ \href {https://doi.org/10.1103/rd84-5g2x} {10.1103/rd84-5g2x}
  (\bibinfo {year} {2026})\BibitemShut {NoStop}%
\bibitem [{\citenamefont {Jafari}\ \emph {et~al.}(2025)\citenamefont {Jafari},
  \citenamefont {Bluhm},\ and\ \citenamefont {DiVincenzo}}]{Jafari2026}%
  \BibitemOpen
  \bibfield  {author} {\bibinfo {author} {\bibfnamefont {S.~A.}\ \bibnamefont
  {Jafari}}, \bibinfo {author} {\bibfnamefont {H.~J.}\ \bibnamefont {Bluhm}},\
  and\ \bibinfo {author} {\bibfnamefont {D.~P.}\ \bibnamefont {DiVincenzo}},\
  }\href {https://arxiv.org/abs/2512.21930} {\bibinfo {title} {Zeeman-like
  coupling to valley degree of freedom in {Si}-based spin qubits}} (\bibinfo
  {year} {2025}),\ \Eprint {https://arxiv.org/abs/arXiv:2512.21930}
  {arXiv:2512.21930} \BibitemShut {NoStop}%
\bibitem [{\citenamefont {Nestoklon}\ \emph {et~al.}(2008)\citenamefont
  {Nestoklon}, \citenamefont {Ivchenko}, \citenamefont {Jancu},\ and\
  \citenamefont {Voisin}}]{Nestoklon2008}%
  \BibitemOpen
  \bibfield  {author} {\bibinfo {author} {\bibfnamefont {M.~O.}\ \bibnamefont
  {Nestoklon}}, \bibinfo {author} {\bibfnamefont {E.~L.}\ \bibnamefont
  {Ivchenko}}, \bibinfo {author} {\bibfnamefont {J.-M.}\ \bibnamefont
  {Jancu}},\ and\ \bibinfo {author} {\bibfnamefont {P.}~\bibnamefont
  {Voisin}},\ }\bibfield  {journal} {\bibinfo  {journal} {Physical Review B}\
  }\textbf {\bibinfo {volume} {77}},\ \href
  {https://doi.org/10.1103/physrevb.77.155328} {10.1103/physrevb.77.155328}
  (\bibinfo {year} {2008})\BibitemShut {NoStop}%
\bibitem [{\citenamefont {Saraiva}\ \emph {et~al.}(2009)\citenamefont
  {Saraiva}, \citenamefont {Calderón}, \citenamefont {Hu}, \citenamefont
  {Das~Sarma},\ and\ \citenamefont {Koiller}}]{Saraiva2009}%
  \BibitemOpen
  \bibfield  {author} {\bibinfo {author} {\bibfnamefont {A.~L.}\ \bibnamefont
  {Saraiva}}, \bibinfo {author} {\bibfnamefont {M.~J.}\ \bibnamefont
  {Calderón}}, \bibinfo {author} {\bibfnamefont {X.}~\bibnamefont {Hu}},
  \bibinfo {author} {\bibfnamefont {S.}~\bibnamefont {Das~Sarma}},\ and\
  \bibinfo {author} {\bibfnamefont {B.}~\bibnamefont {Koiller}},\ }\bibfield
  {journal} {\bibinfo  {journal} {Physical Review B}\ }\textbf {\bibinfo
  {volume} {80}},\ \href {https://doi.org/10.1103/physrevb.80.081305}
  {10.1103/physrevb.80.081305} (\bibinfo {year} {2009})\BibitemShut {NoStop}%
\bibitem [{\citenamefont {Ruskov}\ \emph {et~al.}(2018)\citenamefont {Ruskov},
  \citenamefont {Veldhorst}, \citenamefont {Dzurak},\ and\ \citenamefont
  {Tahan}}]{Ruskov2018}%
  \BibitemOpen
  \bibfield  {author} {\bibinfo {author} {\bibfnamefont {R.}~\bibnamefont
  {Ruskov}}, \bibinfo {author} {\bibfnamefont {M.}~\bibnamefont {Veldhorst}},
  \bibinfo {author} {\bibfnamefont {A.~S.}\ \bibnamefont {Dzurak}},\ and\
  \bibinfo {author} {\bibfnamefont {C.}~\bibnamefont {Tahan}},\ }\bibfield
  {journal} {\bibinfo  {journal} {Physical Review B}\ }\textbf {\bibinfo
  {volume} {98}},\ \href {https://doi.org/10.1103/physrevb.98.245424}
  {10.1103/physrevb.98.245424} (\bibinfo {year} {2018})\BibitemShut {NoStop}%
\bibitem [{\citenamefont {Niquet}\ \emph {et~al.}(2009)\citenamefont {Niquet},
  \citenamefont {Rideau}, \citenamefont {Tavernier}, \citenamefont {Jaouen},\
  and\ \citenamefont {Blase}}]{Niquet2009}%
  \BibitemOpen
  \bibfield  {author} {\bibinfo {author} {\bibfnamefont {Y.~M.}\ \bibnamefont
  {Niquet}}, \bibinfo {author} {\bibfnamefont {D.}~\bibnamefont {Rideau}},
  \bibinfo {author} {\bibfnamefont {C.}~\bibnamefont {Tavernier}}, \bibinfo
  {author} {\bibfnamefont {H.}~\bibnamefont {Jaouen}},\ and\ \bibinfo {author}
  {\bibfnamefont {X.}~\bibnamefont {Blase}},\ }\href
  {https://doi.org/10.1103/PhysRevB.79.245201} {\bibfield  {journal} {\bibinfo
  {journal} {Phys. Rev. B}\ }\textbf {\bibinfo {volume} {79}},\ \bibinfo
  {pages} {245201} (\bibinfo {year} {2009})}\BibitemShut {NoStop}%
\bibitem [{\citenamefont {Guo}\ \emph {et~al.}(2005)\citenamefont {Guo},
  \citenamefont {Yao},\ and\ \citenamefont {Niu}}]{Guo2005}%
  \BibitemOpen
  \bibfield  {author} {\bibinfo {author} {\bibfnamefont {G.~Y.}\ \bibnamefont
  {Guo}}, \bibinfo {author} {\bibfnamefont {Y.}~\bibnamefont {Yao}},\ and\
  \bibinfo {author} {\bibfnamefont {Q.}~\bibnamefont {Niu}},\ }\bibfield
  {journal} {\bibinfo  {journal} {Physical Review Letters}\ }\textbf {\bibinfo
  {volume} {94}},\ \href {https://doi.org/10.1103/physrevlett.94.226601}
  {10.1103/physrevlett.94.226601} (\bibinfo {year} {2005})\BibitemShut
  {NoStop}%
\bibitem [{\citenamefont {Tetlow}\ and\ \citenamefont
  {Gradhand}(2013)}]{Tetlow2013}%
  \BibitemOpen
  \bibfield  {author} {\bibinfo {author} {\bibfnamefont {H.}~\bibnamefont
  {Tetlow}}\ and\ \bibinfo {author} {\bibfnamefont {M.}~\bibnamefont
  {Gradhand}},\ }\bibfield  {journal} {\bibinfo  {journal} {Physical Review B}\
  }\textbf {\bibinfo {volume} {87}},\ \href
  {https://doi.org/10.1103/physrevb.87.075206} {10.1103/physrevb.87.075206}
  (\bibinfo {year} {2013})\BibitemShut {NoStop}%
\bibitem [{\citenamefont {Ezhevskii}\ \emph {et~al.}(2020)\citenamefont
  {Ezhevskii}, \citenamefont {Guseinov}, \citenamefont {Soukhorukov},
  \citenamefont {Novikov}, \citenamefont {Yurasov},\ and\ \citenamefont
  {Gusev}}]{Ezhevskii2020}%
  \BibitemOpen
  \bibfield  {author} {\bibinfo {author} {\bibfnamefont {A.~A.}\ \bibnamefont
  {Ezhevskii}}, \bibinfo {author} {\bibfnamefont {D.~V.}\ \bibnamefont
  {Guseinov}}, \bibinfo {author} {\bibfnamefont {A.~V.}\ \bibnamefont
  {Soukhorukov}}, \bibinfo {author} {\bibfnamefont {A.~V.}\ \bibnamefont
  {Novikov}}, \bibinfo {author} {\bibfnamefont {D.~V.}\ \bibnamefont
  {Yurasov}},\ and\ \bibinfo {author} {\bibfnamefont {N.~S.}\ \bibnamefont
  {Gusev}},\ }\href {https://doi.org/10.1103/PhysRevB.101.195202} {\bibfield
  {journal} {\bibinfo  {journal} {Phys. Rev. B}\ }\textbf {\bibinfo {volume}
  {101}},\ \bibinfo {pages} {195202} (\bibinfo {year} {2020})}\BibitemShut
  {NoStop}%
\bibitem [{\citenamefont {Kawakami}\ \emph {et~al.}(2014)\citenamefont
  {Kawakami}, \citenamefont {Scarlino}, \citenamefont {Ward}, \citenamefont
  {Braakman}, \citenamefont {Savage}, \citenamefont {Lagally}, \citenamefont
  {Friesen}, \citenamefont {Coppersmith}, \citenamefont {Eriksson},\ and\
  \citenamefont {Vandersypen}}]{Kawakami2014}%
  \BibitemOpen
  \bibfield  {author} {\bibinfo {author} {\bibfnamefont {E.}~\bibnamefont
  {Kawakami}}, \bibinfo {author} {\bibfnamefont {P.}~\bibnamefont {Scarlino}},
  \bibinfo {author} {\bibfnamefont {D.~R.}\ \bibnamefont {Ward}}, \bibinfo
  {author} {\bibfnamefont {F.~R.}\ \bibnamefont {Braakman}}, \bibinfo {author}
  {\bibfnamefont {D.~E.}\ \bibnamefont {Savage}}, \bibinfo {author}
  {\bibfnamefont {M.~G.}\ \bibnamefont {Lagally}}, \bibinfo {author}
  {\bibfnamefont {M.}~\bibnamefont {Friesen}}, \bibinfo {author} {\bibfnamefont
  {S.~N.}\ \bibnamefont {Coppersmith}}, \bibinfo {author} {\bibfnamefont
  {M.~A.}\ \bibnamefont {Eriksson}},\ and\ \bibinfo {author} {\bibfnamefont
  {L.~M.~K.}\ \bibnamefont {Vandersypen}},\ }\href
  {https://doi.org/10.1038/nnano.2014.153} {\bibfield  {journal} {\bibinfo
  {journal} {Nature Nanotechnology}\ }\textbf {\bibinfo {volume} {9}},\
  \bibinfo {pages} {666–670} (\bibinfo {year} {2014})}\BibitemShut {NoStop}%
\bibitem [{\citenamefont {Losert}\ \emph {et~al.}(2023)\citenamefont {Losert},
  \citenamefont {Eriksson}, \citenamefont {Joynt}, \citenamefont {Rahman},
  \citenamefont {Scappucci}, \citenamefont {Coppersmith},\ and\ \citenamefont
  {Friesen}}]{Losert2023}%
  \BibitemOpen
  \bibfield  {author} {\bibinfo {author} {\bibfnamefont {M.~P.}\ \bibnamefont
  {Losert}}, \bibinfo {author} {\bibfnamefont {M.~A.}\ \bibnamefont
  {Eriksson}}, \bibinfo {author} {\bibfnamefont {R.}~\bibnamefont {Joynt}},
  \bibinfo {author} {\bibfnamefont {R.}~\bibnamefont {Rahman}}, \bibinfo
  {author} {\bibfnamefont {G.}~\bibnamefont {Scappucci}}, \bibinfo {author}
  {\bibfnamefont {S.~N.}\ \bibnamefont {Coppersmith}},\ and\ \bibinfo {author}
  {\bibfnamefont {M.}~\bibnamefont {Friesen}},\ }\bibfield  {journal} {\bibinfo
   {journal} {Physical Review B}\ }\textbf {\bibinfo {volume} {108}},\ \href
  {https://doi.org/10.1103/physrevb.108.125405} {10.1103/physrevb.108.125405}
  (\bibinfo {year} {2023})\BibitemShut {NoStop}%
\end{thebibliography}%

\pagebreak
\begin{widetext}
\appendix
\section{\label{sec:}Construction of the double group}

The classes of the double space group are
\begin{align*}
C_{1}  =& \{ E \},\\
C_{2}  =& \{ C^{2}_{4x},\, C^{2}_{4y},\, Q C^{2}_{4x},\, Q C^{2}_{4y},\, \\&\mathcal{R} C^{2}_{4x},\, \mathcal{R} C^{2}_{4y},\, \mathcal{R}Q C^{2}_{4x},\, \mathcal{R}Q C^{2}_{4y} \},\\
C_{3}  =& \{ C^{2}_{4z},\, \mathcal{R} C^{2}_{4z} \},\\
C_{4}  =& \{ Q g m_{1},\, g m_{2},\, \mathcal{R}Q g m_{1},\, \mathcal{R} g m_{2} \},\\
C_{5}  =& \{ g S_{4z},\, g S_{4z}^{-1},\, Q g S_{4z},\, Q g S_{4z}^{-1} \},\\
C_{6}  =& \{ g,\, Q g, \mathcal{R} g,\,  \mathcal{R} Q g \},\\
C_{7}  =& \{ g C^{2}_{4x},\, g C^{2}_{4y},\, Q g C^{2}_{4x},\, Q g C^{2}_{4y},\, \\ &\mathcal{R} g C^{2}_{4x},\, \mathcal{R} g C^{2}_{4y},\, \mathcal{R}Q g C^{2}_{4x},\, \mathcal{R}Q g C^{2}_{4y} \},\\
C_{8}  =& \{ g C^{2}_{4z},\, Q g C^{2}_{4z}\},\\
C_{9}  =& \{ m_{1},\, m_{2},\, \mathcal{R} m_{1},\, \mathcal{R} m_{2} \},\\
C_{10} =& \{ S_{4z},\, S_{4z}^{-1},\, Q S_{4z},\, Q S_{4z}^{-1} \},\\
C_{11} =& \{ Q m_{1},\, Q m_{2},\, \mathcal{R}Q m_{1},\, \mathcal{R}Q m_{2} \},\\
C_{12} =& \{ Q g m_{2},\, g m_{1},\, \mathcal{R}Q g m_{2},\, \mathcal{R} g m_{1} \},\\
C_{13} =& \{ Q C^{2}_{4z},\, \mathcal{R}Q C^{2}_{4z} \},\\
C_{14} =& \{ Q \},\\
C_{15} =& \{ \mathcal{R} E \},\\
C_{16} =& \{ \mathcal{R} g S_{4z},\, \mathcal{R} g S_{4z}^{-1},\, \mathcal{R}Q g S_{4z},\, \mathcal{R}Q g S_{4z}^{-1} \},\\
C_{17} =& \{\mathcal{R} g C^{2}_{4z},\, \mathcal{R}Q g C^{2}_{4z} \},\\
C_{18} =& \{ \mathcal{R} S_{4z},\, \mathcal{R} S_{4z}^{-1},\, \mathcal{R}Q S_{4z},\, \mathcal{R}Q S_{4z}^{-1} \},\\
C_{19} =& \{ \mathcal{R}Q \}.
\end{align*}
There are 64 objects in 19 classes. This means that the dimensions $l_i$ of the irreducible representations must fulfill
\begin{equation}
    \sum_{i=1}^{19}l_i^2 = 64.
\end{equation}
This equation has eight distinct solutions. Imposing the additional requirement that all conjugacy classes of the single group must also appear in the double group — either unchanged or multiplied by $\{E, \mathcal{R}\}$ — reduces this number to three. These "extra" representations (that do not exist in single group) satisfy
\begin{equation}
\sum_{i=1}^{5} l_i^2 = 32
\label{eq:dims_IRs}
\end{equation}
since the dimensions of the first 14 irreducible representations are fixed by the single group, which contains 32 elements. The solutions for the dimensions of the extra representations are $(2,2,2,2,4)$, $(1,1,1,2,5)$, and $(1,2,3,3,3)$. Only the first set of solutions contains a spinor representation of dimension $2$ and a spinor-valleyor representation of dimension $2\times 2=4$. The last two sets containing four \emph{odd}-dimensional representations can be refuted yet on another ground: Trying to construct a character table with those dimensions leads to contradiction. 

To construct the character table of the double group, we begin by extending the single group’s character table (cf. Ref.~\cite{Jafari2026}). The characters for classes 15 through 19 are taken to be identical to those of classes 1, 5, 8, 10, and 14, respectively, since these classes differ only by multiplication with $\mathcal{R}$ and the corresponding representations originate from the spinless single group where $\mathcal{R}\equiv E$. We then introduce five additional irreducible representations, denoted $Y_1$ through $Y_5$, which arise from incorporating spin into the group.

The two-dimensional representation $Y_1$ corresponds to the spinor representation, and its characters are obtained from the traces of the corresponding class operators in spin space. Notably, the character vanishes for any class containing both $C$ and $\mathcal{R}C$, as these correspond to $C$ and $-C$ in spin space that is consistent only with zero solution. In contrast to the spinless representations of the single group, the spinor representations yield characters for classes 15 through 19 that are the negatives of those for classes 1, 5, 8, 10, and 14, reflecting that $\mathcal{R} = -E$ in spin space.

Among the three candidate solutions to Eq.~\eqref{eq:dims_IRs}, only one results in a character table that satisfies the orthogonality relations while remaining consistent with the structure of the single group. Although the other two solutions formally allow the construction of character tables, they fail to meet this additional constraint. This uniquely determines the character table shown in Table~\ref{tab:double_group_char_table}. The IR of some common objects is given.

\begin{table*}[t]
\centering
\renewcommand{\arraystretch}{1.1} 
\setlength{\tabcolsep}{3pt}       
\begin{tabular}{|c|*{5}{c}||*{5}{c}||*{4}{c}||*{5}{c}||l|}
\hline
 & $\mathcal{C}_{1}$ & $\mathcal{C}_{2}$ & $\mathcal{C}_{3}$ & $\mathcal{C}_{4}$ & $\mathcal{C}_{5}$ & $\mathcal{C}_{6}$ & $\mathcal{C}_{7}$ & $\mathcal{C}_{8}$ & $\mathcal{C}_{9}$ & $\mathcal{C}_{10}$ & $\mathcal{C}_{11}$ & $\mathcal{C}_{12}$ & $\mathcal{C}_{13}$ & $\mathcal{C}_{14}$ & $\mathcal{C}_{15}$ & $\mathcal{C}_{16}$ & $\mathcal{C}_{17}$ & $\mathcal{C}_{18}$ & $\mathcal{C}_{19}$ & Objects \\
\hline
$M_{1}$     & 1 & 1 & 1 & 1 & 1 & 1 & 1 & 1 & 1 & 1 & 1 & 1 & 1 & 1 & 1 & 1 & 1 & 1 & 1 & $\tau_0, \sigma_0, k_z^2, k_x^2+k_y^2, \varepsilon_{zz}, \varepsilon_{xx}+\varepsilon_{yy}$ \\
$M_{2}$     & 1 & 1 & 1 & -1 & -1 & 1 & 1 & 1 & -1 & -1 & -1 & -1 & 1 & 1 & 1 & -1 & 1 & -1 & 1 & $k_x^2-k_y^2, \varepsilon_{xx}-\varepsilon_{yy}$\\
$M_{3}$     & 1 & -1 & 1 & -1 & 1 & 1 & -1 & 1 & -1 & 1 & -1 & -1 & 1 & 1 & 1 & 1 & 1 & 1 & 1 & $R_z, \sigma_z$ \\
$M_{4}$     & 1 & -1 & 1 & 1 & -1 & 1 & -1 & 1 & 1 & -1 & 1 & 1 & 1 & 1 & 1 & -1 & 1 & -1 & 1 & $\tau_2, k_xk_y, \varepsilon_{xy}$ \\
$M_{5}$     & 2 & 0 & -2 & 0 & 0 & 2 & 0 & -2 & 0 & 0 & 0 & 0 & -2 & 2 & 2 & 0 & -2 & 0 & 2 & $(k_x,k_y), (P_x,P_y)$\\
\hline \hline
$M'_{1}$    & 1 & 1 & 1 & 1 & 1 & -1 & -1 & -1 & -1 & -1 & -1 & 1 & 1 & 1 & 1 & 1 & -1 & -1 & 1 & $P_xR_x + P_yR_y$\\
$M'_{2}$    & 1 & 1 & 1 & -1 & -1 & -1 & -1 & -1 & 1 & 1 & 1 & -1 & 1 & 1 & 1 & -1 & -1 & 1 & 1 & $\tau_1, P_xR_x-P_yR_y$\\
$M'_{3}$    & 1 & -1 & 1 & -1 & 1 & -1 & 1 & -1 & 1 & -1 & 1 & -1 & 1 & 1 & 1 & 1 & -1 & -1 & 1 & $\tau_3, k_z, P_xR_y-P_yR_x$\\
$M'_{4}$    & 1 & -1 & 1 & 1 & -1 & -1 & 1 & -1 & -1 & 1 & -1 & 1 & 1 & 1 & 1 & -1 & -1 & 1 & 1 & $P_xR_y+P_yR_x$ \\
$M'_{5}$    & 2 & 0 & -2 & 0 & 0 & -2 & 0 & 2 & 0 & 0 & 0 & 0 & -2 & 2 & 2 & 0 & 2 & 0 & 2 & $(R_x, R_y), (\varepsilon_{yz, \varepsilon_{xz}}), (\sigma_x, \sigma_y)$ \\
\hline \hline
$X_{1}$     & 2 & 0 & 2 & 0 & 0 & 0 & 0 & 0 & 2 & 0 & -2 & 0 & -2 & -2 & 2 & 0 & 0 & 0 & -2 & $(X_1^+, X_1^-)$\\
$X_{2}$     & 2 & 0 & 2 & 0 & 0 & 0 & 0 & 0 & -2 & 0 & 2 & 0 & -2 & -2 & 2 & 0 & 0 & 0 & -2 &\\
$X_{3}$     & 2 & 0 & -2 & 2 & 0 & 0 & 0 & 0 & 0 & 0 & 0 & -2 & 2 & -2 & 2 & 0 & 0 & 0 & -2 &\\
$X_{4}$     & 2 & 0 & -2 & -2 & 0 & 0 & 0 & 0 & 0 & 0 & 0 & 2 & 2 & -2 & 2 & 0 & 0 & 0 & -2 &\\
\hline \hline
$Y_{1}$     &  2 & 0 & 0 & 0 & -$\sqrt{2}$ & 0 & 0 & -2 & 0 & -$\sqrt{2}$ & 0 & 0 & 0 & 2 & -2 & $\sqrt{2}$ & 2 & $\sqrt{2}$ & -2 & $(\chi^+, \chi^-)$ \\
$Y_{2}$     &  2 & 0 & 0 & 0 & $\sqrt{2}$ & 0 & 0 & -2 & 0 & $\sqrt{2}$ & 0 & 0 & 0 & 2 & -2 & -$\sqrt{2}$ & 2 & -$\sqrt{2}$ & -2 &\\
$Y_{3}$     &  2 & 0 & 0 & 0 & -$\sqrt{2}$ & 0 & 0 & 2 & 0 & $\sqrt{2}$ & 0 & 0 & 0 & 2 & -2 & $\sqrt{2}$ & -2 & -$\sqrt{2}$ & -2 &\\
$Y_{4}$     &  2 & 0 & 0 & 0 & $\sqrt{2}$ & 0 & 0 & 2 & 0 & -$\sqrt{2}$ & 0 & 0 & 0 & 2 & -2 & -$\sqrt{2}$ & -2 & $\sqrt{2}$ & -2 &\\
$Y_{5}$     &  4 & 0 & 0 & 0 & 0 & 0 & 0 & 0 & 0 & 0 & 0 & 0 & 0 & -4 &  -4 & 0 & 0 & 0 & 4 & $(\chi^+, \chi^-)\otimes (X_1^+, X_1^-)$\\
\hline
\end{tabular}
\caption{Character table of the double group. The functions $X_1^\pm$ that belong to IR $X_1$ are two Bloch functions forming a basis for valleyors, whereas $\chi^\pm$ that belong to IR $Y_1=D_{1/2}$ are spin states forming a basis for spinors. The IR $Y_5=Y_1\otimes X_1$ contains all the transformation properties of  the spin-valley basis.}
\label{tab:double_group_char_table}
\end{table*}

To ensure the invariance of the Hamiltonian under the group action, any permissible term in $H$ must belong to the scalar IR $M_1$. To compute such scalar terms, it is useful to construct a multiplication table. Table~\ref{tab:multiplication_table} shows how the product of any two IRs decomposes into other IRs. For example, the product of $M_2$ and $M'_3$ belongs to $M'_4$. Higher-dimensional IRs decompose into a direct sum of irreducible components. For example, $M_5\times M_5 = M_1 + M_2 + M_3 + M_4$. This decomposition is reflected in the fact that two polar vectors $(P_x, P_y)$ and $(k_x, k_y)$ can be combined into symmetry-adapted bilinears transforming as the different irreducible representations: $k_x P_x + k_y P_y$ ($M_1$), $k_x P_x - k_y P_y$ ($M_2$), $k_x P_y - k_y P_x$ ($M_3$), and $k_x P_y + k_y P_x$ ($M_4$).

\begin{table*}[h]
\centering
\scriptsize
\renewcommand{\arraystretch}{1.1}
\setlength{\tabcolsep}{2pt}
\resizebox{\textwidth}{!}{
\begin{tabular}{|c|c|c|c|c|c|c|c|c|c|c|c|c|c|c|c|c|c|c|c|}
\hline
 & $M_1$ & $M_2$ & $M_3$ & $M_4$ & $M_5$ & $M'_1$ & $M'_2$ & $M'_3$ & $M'_4$ & $M'_5$ & $X_1$ & $X_2$ & $X_3$ & $X_4$ & $Y_1$ & $Y_2$ & $Y_3$ & $Y_4$ & $Y_5$ \\
\hline
$M_1$ & $M_1$ &  &  &  &  &  &  &  &  &  &  &  &  &  &  &  &  &  &  \\
\hline
$M_2$ & $M_2$ & $M_1$ &  &  &  &  &  &  &  &  &  &  &  &  &  &  &  &  & \\
\hline
$M_3$ & $M_3$ & $M_4$ & $M_1$ &  &  &  &  &  &  &  &  &  &  &  &  &  &  &  &  \\
\hline
$M_4$ & $M_4$ & $M_3$ & $M_2$ & $M_1$ &  &  &  &  &  &  &  &  &  &  &  &  &  &  & \\
\hline
$M_5$ & $M_5$ & $M_5$ & $M_5$ & $M_5$ & $M_1$+$M_2$&  &  &  &  &  &  &  &  &  &  &  &  &  &  \\
&&&&& +$M_3$+$M_4$ &  &  &  &  &  &  &  &  &  &  &  &  &  &  \\

\hline
$M'_1$ & $M'_1$ & $M'_2$ & $M'_3$ & $M'_4$ & $M'_5$ & $M_1$ &  &  &  &  &  &  &  &  &  &  &  &  &  \\
\hline
$M'_2$ & $M'_2$ & $M'_1$ & $M'_4$ & $M'_3$ & $M'_5$ & $M_2$ & $M_1$ &  &  &  &  &  &  &  &  &  &  &  &  \\
\hline
$M'_3$ & $M'_3$ & $M'_4$ & $M'_1$ & $M'_2$ & $M'_5$ & $M_3$ & $M_4$ & $M_1$ &  &  &  &  &  &  &  &  &  &  &  \\
\hline
$M'_4$ & $M'_4$ & $M'_3$ & $M'_2$ & $M'_1$ & $M'_5$ & $M_4$ & $M_3$ & $M_2$ & $M_1$ &  &  &  &  &  &  &  &  &  &  \\
\hline
$M'_5$ & $M'_5$ & $M'_5$ & $M'_5$ & $M'_5$ & $M'_1$+$M'_2$ & $M_5$ & $M_5$ & $M_5$ & $M_5$ & $M_1$+$M_2$  &  &  &  &  &  &  &  &  &  \\
&&&&&+$M'_3$+$M'_4$ &&&&&+$M_3$+$M_4$ &  &  &  &  &  &  &  &  &  \\
\hline
$X_1$ & $X_1$ & $X_2$ & $X_2$ & $X_1$ & $X_3$+$X_4$ & $X_2$ & $X_1$ & $X_1$ & $X_2$ & $X_3$+$X_4$ & $M_1$+$M_4$ &  &  &  &  &  &  &  &  \\
&&&&&&&&&&& +$M'_2$+$M'_3$ &  &  &  &  &  &  &  &   \\

\hline
$X_2$ & $X_2$ & $X_1$ & $X_1$ & $X_2$ & $X_3$+$X_4$ & $X_1$ & $X_2$ & $X_2$ & $X_1$ & $X_3$+$X_4$ & $M_2$+$M_3$ & $M_1$+$M_4$ &  &  &  &  &  & &   \\
&&&&&&&&&&&+$M'_1$+$M'_4$ & +$M'_2$+$M'_3$ &  &  &  &  &  &  & \\

\hline
$X_3$ & $X_3$ & $X_4$ & $X_4$ & $X_3$ & $X_1$+$X_2$ & $X_3$ & $X_4$ & $X_4$ & $X_3$ & $X_1$+$X_2$ & $M_5$+$M'_5$ & $M_5$+$M'_5$ & $M_1$+$M_4$ &  &  &  &  &  &  \\
&&&&&&&&&&&&& +$M'_1$+$M'_4$&  &  &  &  & &  \\

\hline
$X_4$ & $X_4$ & $X_3$ & $X_3$ & $X_4$ & $X_1$+$X_2$ & $X_4$ & $X_3$ & $X_3$ & $X_4$ & $X_1$+$X_2$ & $M_5$+$M'_5$ & $M_5$+$M'_5$ & $M_2$+$M_3$& $M_1$+$M_4$  &  &  &  & & \\
&&&&&&&&&&&&& +$M'_2$+$M'_3$ & +$M'_1$+$M'_4$&  &  &  &  &  \\

\hline
$Y_1$ & $Y_1$ & $Y_2$ & $Y_1$ & $Y_2$ & $Y_3$+$Y_4$ & $Y_3$ & $Y_4$ & $Y_3$ & $Y_4$ & $Y_1$+$Y_2$ & $Y_5$ & $Y_5$ & $Y_5$ & $Y_5$ & $M_1$+$M_3$  &  &  &&   \\
&&&&&&&&&&&&&&&+$M'_5$&  &  &  &  \\

\hline
$Y_2$ & $Y_2$ & $Y_1$ & $Y_2$ & $Y_1$ & $Y_3$+$Y_4$ & $Y_4$ & $Y_3$ & $Y_4$ & $Y_3$ & $Y_1$+$Y_2$ & $Y_5$ & $Y_5$ & $Y_5$ & $Y_5$ & $M_2$+$M_4$  & $M_1$+$M_3$ &  &  &  \\
&&&&&&&&&&&&&&&+$M'_5$& +$M'_5$ &  &  &  \\

\hline
$Y_3$ & $Y_3$ & $Y_4$ & $Y_3$ & $Y_4$ & $Y_1$+$Y_2$ & $Y_1$ & $Y_2$ & $Y_1$ & $Y_2$ & $Y_3$+$Y_4$ & $Y_5$ & $Y_5$ & $Y_5$ & $Y_5$ & $M_5$+$M'_1$ & $M'_5$+$M'_2$ & $M_1$+$M_3$ &  &  \\
&&&&&&&&&&&&&&& +$M'_3$& +$M'_4$& +$M'_5$ & &   \\

\hline
$Y_4$ & $Y_4$ & $Y_3$ & $Y_4$ & $Y_3$ & $Y_1$+$Y_2$ & $Y_2$ & $Y_1$ & $Y_2$ & $Y_1$ & $Y_3$+$Y_4$ & $Y_5$ & $Y_5$ & $Y_5$ & $Y_5$ & $M_5$+$M'_2$ & $M_5$+$M'_1$& $M_2$+$M_4$ & $M_1$+$M_3$  &  \\
&&&&&&&&&&&&&&& +$M'_4$ & +$M'_3$ & +$M'_5$&+$M'_5$& \\

\hline
$Y_5$ & $Y_5$ & $Y_5$ & $Y_5$ & $Y_5$ & $2Y_5$ & $Y_5$ & $Y_5$ & $Y_5$ & $Y_5$ & $2Y_5$ & $Y_1$+$Y_2$ & $Y_1$+$Y_2$ & $Y_1$+$Y_2$ & $Y_1$+$Y_2$ & $X_1$+$X_2$ & $X_1$+$X_2$ & $X_1$+$X_2$ & $X_1$+$X_2$ & $M_1$+$M_2$ \\
&&&&&&&&&&& +$Y_3$+$Y_4$ &+$Y_3$+$Y_4$ &+$Y_3$+$Y_4$ &+$Y_3$+$Y_4$ &+$X_3$+$X_4$ &+$X_3$+$X_4$ &+$X_3$+$X_4$ &+$X_3$+$X_4$ &+$M_3$+$M_4$ \\
&&&&&&&&&&&&&&&&&&& +$2M_5$+$M'_1$\\
&&&&&&&&&&&&&&&&&&& +$M'_2$+$M'_3$\\
&&&&&&&&&&&&&&&&&&& +$M'_4$+$2M'_5$\\

\hline
\end{tabular}
}
\caption{Multiplication table of the IRs of the double space group}
\label{tab:multiplication_table}
\end{table*}

\section{Spin-orbit-valley terms}
To construct all spin-orbit terms allowed in the Hamiltonian, we use the method of invariants: any term appearing in $H$ must be invariant under the group action, and hence must belong to the trivial (scalar) representation $M_1$. The momentum doublet $(k_x,k_y)$ transforms as $M_5$ and the spin doublet $(\sigma_x,\sigma_y)$ as $M'_5$. From Table~\ref{tab:multiplication_table}, $M_5\times M'_5 = M'_1+M'_2+M'_3+M'_4$, so the four bilinears (tensorial spin-orbit objects) $k_x\sigma_x+k_y\sigma_y$, $k_x\sigma_x-k_y\sigma_y$, $k_x\sigma_y-k_y\sigma_x$, and $k_x\sigma_y+k_y\sigma_x$ belong to $M'_1$, $M'_2$, $M'_3$, and $M'_4$, respectively.

Each bilinear becomes an invariant spin-orbit-valley term only once multiplied by a valley matrix $\tau_i$ of the same IR, since $M'_j\times M'_j=M_1$. Among the four valley matrices, only $\tau_1$ ($M'_2$) and $\tau_3$ ($M'_3$) match one of the bilinears, leaving two candidate terms in the absence of external fields:
\begin{equation}
(k_x\sigma_x-k_y\sigma_y)\,\tau_1 \quad\text{and}\quad (k_x\sigma_y-k_y\sigma_x)\,\tau_3.
\end{equation}
Because $\tau_3$ is odd under time reversal while $\tau_1$ and $\tau_2$ are even, only the Dresselhaus SOC $(k_x\sigma_x-k_y\sigma_y)\,\tau_1$ survives as an intrinsic bulk term.

\begin{table}[b]
\centering
\renewcommand{\arraystretch}{1.3}
\begin{tabular}{|c|}
\hline
$(k_x\sigma_x-k_y\sigma_y)\,\tau_1$ \\
$(k_x\sigma_x-k_y\sigma_y)\,E_z\,\tau_2$ \\
$(k_x\sigma_y-k_y\sigma_x)\,E_z\,\tau_0$ \\
$(k_x\sigma_x+k_y\sigma_y)\,B_z\,\tau_3$ \\
\hline
\end{tabular}
\caption{Symmetry-allowed TR-even spin-orbit-valley terms up to linear order in $E_z$ and $B_z$.}
\label{tab:so_terms}
\end{table}

We can also consider first-order external fields: an electric field $E_z$, which belongs to $M'_3$ (similar to like $k_z$), and a magnetic field $B_z$, which transforms as $M_3$ like $R_z,\sigma_z$. Note that $E_z$ is TR-even, and $B_z$ is TR-odd. Combined with $E_z$, the $M'_2$ bilinear now needs a valley matrix of IR $M'_2\times M'_3=M_4$, the IR which contains $\tau_2$:
\begin{equation}
(k_x\sigma_x-k_y\sigma_y)\,E_z\,\tau_2, \qquad M'_2\times M'_3\times M_4=M_1.
\end{equation}
The $M'_3$ bilinear combined with $E_z$ instead needs $M'_3\times M'_3=M_1$, i.e.\ $\tau_0$:
\begin{equation}
(k_x\sigma_y-k_y\sigma_x)\,E_z\,\tau_0, \qquad M'_3\times M'_3\times M_1=M_1.
\end{equation}
With $B_z$, the $M'_1$ bilinear requires $M'_1\times M_3=M'_3$, i.e.\ $\tau_3$:
\begin{equation}
(k_x\sigma_x+k_y\sigma_y)\,B_z\,\tau_3, \qquad M'_1\times M_3\times M'_3=M_1.
\end{equation}
The remaining possibility, $M'_4\times M_3=M'_2$, would give a term $(k_x\sigma_y + k_y\sigma_x)B_z\tau_1$, but this term is again forbidden by time-reversal symmetry, just as in the field-free case. Together with the intrinsic Dresselhaus term, this gives the complete set of SOV couplings linear in $E_z$ and $B_z$.
The resulting terms are summarized in table~\ref{tab:so_terms}.

\section{Micromagnet calculations}
To describe the difference in Zeeman splitting between the valley eigenstates observed in Ref.~\cite{Ferdous2018}, we consider the Hamiltonian
\begin{equation}
    H = H_0 + H_Z + H_{\Delta B}.
\end{equation}
Here, $H_0$ denotes any terms diagonal in spin space, which are not relevant for the following discussion, $H_Z$ is the Zeeman Hamiltonian $H_Z = \frac{1}{2}g\mu\vec{B}\cdot\vec{\sigma}$, and $H_{\Delta B}$ describes the contribution of the inhomogeneity of the $B$-field. The contribution of the inhomogeneous part $H_{\Delta B}$ on the Zeeman splittings of each valley state is negligible against the effect of the homogeneous $B$-field, but becomes relevant when computing the difference in Zeeman splitting between the two valley eigenstates. To capture such terms one needs to identify the IR of the objects of the form $\partial_lB_m$ and then look into their multiplication with $\sigma_i\tau_j$ objects and pick the TR-invariant combinations. 

Let us first consider the invariant 
\begin{equation}
    H_{\Delta B}^1 = \lambda_1( \partial_x B_z \sigma_y + \partial_y B_z \sigma_x )\tau_1.
\end{equation}
The eigenvalues of $H_Z + H_{\Delta B}^1$ are
\begin{align}
    E_{\eta, s} &= s\sqrt{\left(\frac{1}{2}g\mu B_x + \eta\lambda_1\frac{dB_z}{dy}\right)^2 + \left(\frac{1}{2}g\mu B_y + \eta\lambda_1\frac{dB_z}{dx}\right)^2 + \left(\frac{1}{2}g\mu B_z\right)^2}\\
    &= \frac{sg\mu}{2}\sqrt{\left( B_x + 2\eta\frac{\lambda_1}{g\mu}\frac{dB_z}{dy}\right)^2 + \left( B_y + 2\eta\frac{\lambda_1}{g\mu}\frac{dB_z}{dx}\right)^2 + B_z^2},
\end{align}
where $s = \pm$ denotes the spin eigenstates and $\eta = \pm$ the two $\tau_1$ valley eigenstates. The difference in Zeeman splitting is given by
\begin{equation}
    \Delta_\tau E_Z^1 = (E^1_{+,\uparrow}-E^1_{+,\downarrow}) - (E^1_{-,\uparrow}-E^1_{-,\downarrow}) = 2(E^1_{+,\uparrow}-E^1_{-,\uparrow}),
\end{equation}
where $\Delta_\tau$ emphasizes the difference between the Zeeman splitting of the two valleys. 
Treating $H_{\Delta B}^1$ as a perturbation, we can approximate
\begin{equation}
    E_{\eta, s} \approx \frac{sg\mu |\vec{B}|}{2}\sqrt{1 +  4\eta \frac{\lambda_1}{g\mu |\vec{B}|^2}\left(B_x\frac{dB_z}{dy} + B_y\frac{dB_z}{dx}\right) }.
\end{equation}
It follows that 
\begin{align}
    \Delta_\tau E_Z^1 &\approx g\mu |\vec{B}|\left(\sqrt{1 +  4 \frac{\lambda_1}{g\mu |\vec{B}|^2}\left(B_x\frac{dB_z}{dy} + B_y\frac{dB_z}{dx}\right)} - \sqrt{1 -  4 \frac{\lambda_1}{g\mu |\vec{B}|^2}\left(B_x\frac{dB_z}{dy} + B_y\frac{dB_z}{dx}\right)}\right) \\
    &\approx 4 \frac{\lambda_1}{|\vec{B}|}\left(B_x\frac{dB_z}{dy} + B_y\frac{dB_z}{dx}\right),
    \label{eq:Delta_E_Z_1}
\end{align}
where we have used $\sqrt{1+x}\approx x/2$ for small $x$.

The calculation for the other invariant term
\begin{equation}
    H_{\Delta B}^2 = \lambda_2( \partial_x B_y  + \partial_y B_x )\sigma_z\tau_1
\end{equation}
is analogous and yields the Zeeman splitting difference
\begin{equation}
    \Delta_\tau E_Z^2 \approx 4\lambda_2 \frac{B_z}{|\vec{B}|}\left(\frac{dB_y}{dx} + \frac{dB_x}{dy}\right).
\end{equation}

To approximate $\lambda_1$, we use the values of $d\vec{B}/d\vec{r}$ at $\theta=0^\circ$ and $\theta=90^\circ$ from the supplementary material to Ref.~\cite{Ferdous2018}. Note that the axes $x', y'$ in Ref.~\cite{Ferdous2018} are turned by $45^\circ$, such that we must do a coordinate transformation to use the values provided. In these coordinates, and using the approximation $B_{x'} = \sin \theta |\vec{B}_{\text{ext}}|$, $B_{y'} = \cos \theta |\vec{B}_{\text{ext}}|$, Eq. \ref{eq:Delta_E_Z_1} can be written as
\begin{equation}
    \Delta_\tau E_Z^1 \approx 4\lambda_1\left(-\sin\theta dB_z/dx' + \cos\theta dB_z/dy'\right).
\end{equation}
At $\theta=0^\circ$, $dB_z/dy' = -0.052$mT/nm, and t $\theta=90^\circ$, $dB_z/dx' = -0.456$mT/nm. Using Fig. 1c from Ref.~\cite{Ferdous2018}, we can approximate
\begin{align}
    &\left.h(f_{v-}-f_{v+})\right|^{\theta=90^\circ}_{0^\circ} \approx \left.\Delta E_Z^1\right|^{\theta=90^\circ}_{0^\circ} \\
    \Rightarrow \quad & h\cdot 6.5\text{MHz} \approx 4\lambda_1\cdot 0.508 \text{mT/nm} \\
    \Rightarrow \quad & \lambda_1 \approx 13 \mu\text{eV nm/T}.
\end{align}

\end{widetext}

\end{document}